\documentclass[preprint,prd,nofootinbib]{revtex4}

\usepackage{graphicx}  
\usepackage{dcolumn}   
\usepackage{bm}        
\usepackage{amssymb}   

\usepackage{color} 

\usepackage{amsmath}
\usepackage{graphicx}

\usepackage{floatrow}

\usepackage{float}

\usepackage{wrapfig}

\usepackage{hyperref}

\begin{document}




\title{Interaction between gravitational radiation and electromagnetic 
radiation}

\author{Preston Jones}
\thanks{Preston.Jones1@erau.edu}
\affiliation{Embry Riddle Aeronautical University, Prescott, AZ 86301}
\author{Douglas Singleton}
\thanks{dougs@mail.fresnostate.edu}
\affiliation{California State University Fresno, Fresno, CA 93740}

\date{\today}

\begin{abstract}
In this review paper we investigate the connection between 
gravity and electromagnetism from Faraday to the present day. The particular 
focus is on the connection between gravitational and electromagnetic 
radiation. We discuss electromagnetic radiation produced when a gravitational 
wave passes through a magnetic field. We then discuss the interaction of 
electromagnetic radiation with gravitational waves via Feynman diagrams 
of the process $graviton + graviton \to photon + photon$. Finally we review 
recent work on the vacuum production of counterpart electromagnetic radiation by 
gravitational waves.
\end{abstract}

\maketitle

\section{Historical Introduction}

The late 1500's and early 1600's were a remarkable period in the evolution of human thought. It might be reasonably argued that during this period Galileo Galilee put into practice the modern scientific method for describing and understanding natural processes. An equally important advancement in our way of thinking about the world was an emerging conviction of the universality of causes. This extraordinary new way of understanding the world around us is often associated with a slightly later period and with Isaac Newton. The notion that the laws of nature applied equally everywhere was indeed imagined in this earlier period by Johannes Kepler. In particular Kepler proposed that the principles that governed the movement of the planets was the same as on Earth. Kepler's thinking of a universal nature of physical properties both celestial and terrestrial is evident in his own words \cite{Holton88} : ``I am occupied with the investigation of the physical causes. My aim in this is to show that the celestial machine is to be likened not to a divine organism but rather to a clockwork ..., insofar as nearly all the manifold movements are carried out by means of a single, quite simple magnetic force, as in the case of a clockwork all motion are caused by a simple weight. Moreover, I show how this physical conception is to be presented through calculation and geometry." Kepler's way of thinking about the motions of the planets and the universality of the laws of physics would be completely recognizable to every modern physicist.

\begin{figure}[H]
\centering
\includegraphics[width=75mm]{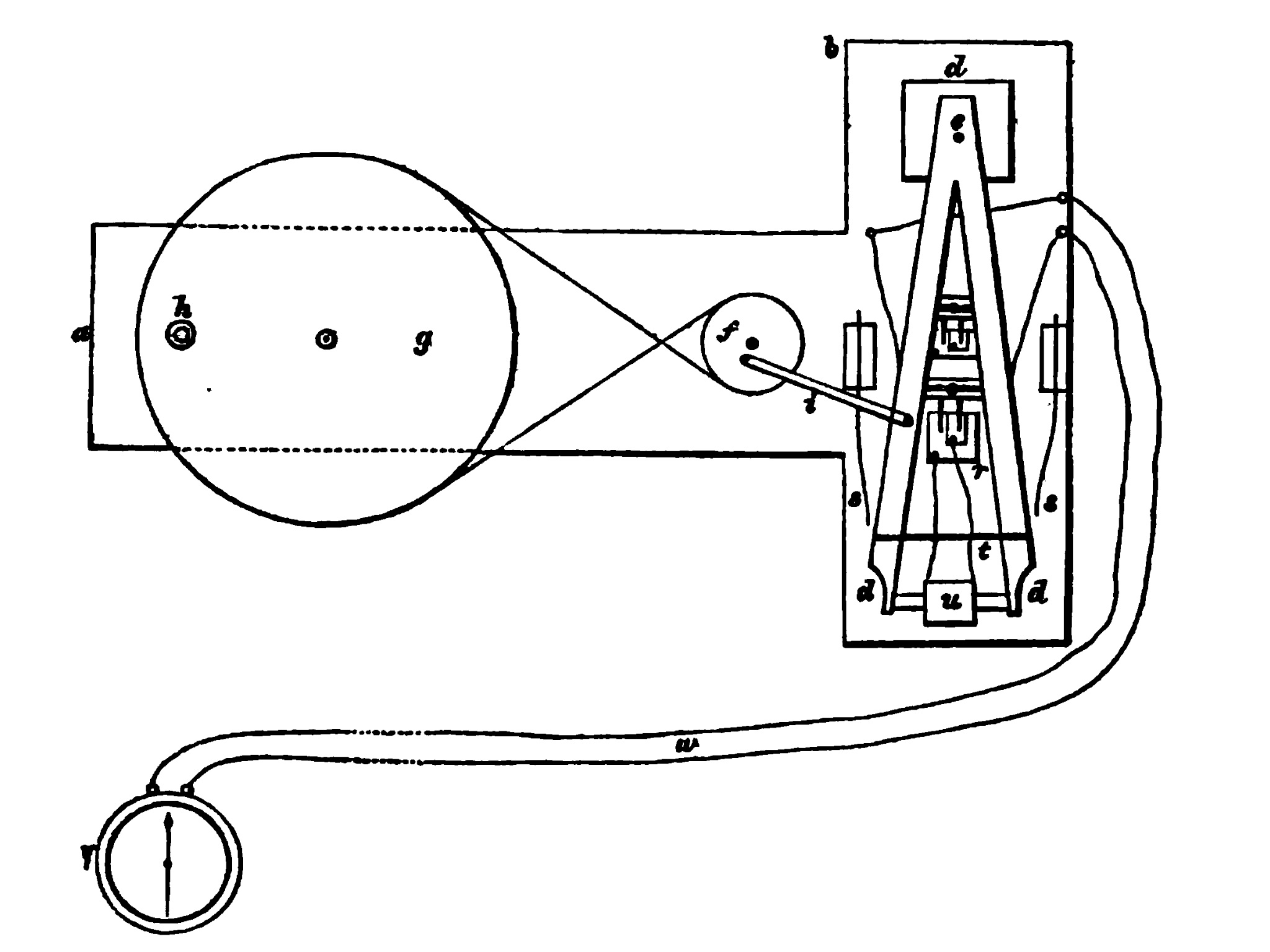}
\caption{This figure is from Faraday's laboratory notebook describing the apparatus he constructed to conduct experiments on the relationship between gravity and electromagnetic induction \cite{Faraday1885}. \label{FaradayFig}}
\end{figure}

While Kepler's conviction of the relationship between the motion of the planets and processes on Earth helped inspire our modern way of thinking, he was of course mistaken in making the association between gravity and a ``simple magnetic force''. However, even this mistake was an inspired effort to describe the world around us in terms of physical causes. The universality of physical principles quickly became a central theme in the development of physics and an inspiration for Newton and those that followed. This expectation of the universality of celestial and terrestrial processes and Kepler's expectation of  universality in a connection between magnetism and the motion of the planets is evident in Faraday's experimental investigations. Some time around the 1850's Faraday conducted experiments to demonstrate the possible connection between the gravitational field and the electromagnetic field. Faraday constructed an experimental apparatus in an effort to measure the magnitude of electromagnetic induction associated with a gravitational field as shown in Fig. \ref{FaradayFig}. Faraday's results failed to demonstrate any relation between gravity and electricity but his commitment to this idea of universality was unwavering, \cite{Faraday1885}, ``Here end my trials for the present. The results are negative. They do not shake my strong feeling of the existence of a relation between gravity and electricity, though they give no proof that such a relation exists."

While Faraday's experiments were not successful, later theoretical research by Skobelev \cite{Skobelev75} in 1975 supported Faraday's ``strong feeling" by demonstrating a non-zero amplitude for the interaction of gravitons and photons in both scattering and annihilation. This association between gravity and electromagnetism was also described around the same time by Gibbons \cite{gibbons} in noting that, ``Indeed since a `graviton' presumably in some sense carries light-like momentum the creation of one or more particles with time-like or light-like momentum would violate the conservation of momentum unless the created particles were massless and precisely aligned with the momentum of the graviton". These kinematic restrictions for conversion of massless particles have also been studied more recently and in greater detail by Fiore and Modanese \cite{Fiore96,Modanese95}. The processes of graviton and photon interaction described by Skobelev and Gibbons are exceedingly small \cite{Skobelev75} but are non-zero. Our more recent research has expanded on this interaction of gravity/gravitons and electromagnetism/photons through annihilation and scattering processes by recognizing the contribution of the external gravitational field associated with a gravitational wave \cite{Jones15,Jones16,Jones17,Jones18,Gretarsson18}. Our study of the vacuum production of light by a gravitational wave differs from Skobelev in that the amplitudes of the ``tree level diagrams" would be dependent on the strength of the external gravitational field or equally the strain amplitude of the gravitational wave. This type of semi-classical conversion process between gravitational and electromagnetic fields was described more broadly by Davies \cite{Davies01} ``One result is that rapidly changing gravitational fields can create particles from the vacuum, and in turn the back-reaction on the gravitational dynamics operates like a damping force." The back-reaction on the gravitational wave was shown to be small compared to the gravitational wave luminosity but sufficient to be detectable under the right circumstances \cite{Jones17}. 

In this brief review we will specifically outline the relationship between gravity and electricity for the special case of gravitational and electromagnetic radiation. While we take a historical perspective leading to current research no effort will be made to present the historical formalism. Instead we will present the ideas relating the association between gravitational and electromagnetic radiation and in particular the vacuum production of electromagnetic radiation by a gravitational wave using modern notation and mathematical formalism.

\section{Electromagnetism and light}

A general review of the research on the relationship between gravity and electricity would completely preclude any possibility of being brief. We will instead focus our attention on the radiation regimes. The current understanding of the radiation regime for electricity began with James Clerk Maxwell's modification of Ampere's law \cite{Maxwell2} to include the displacement current. This modification led to a wave equation solution to the equations of electromagnetism. Maxwell immediately recognized this wave equation as a description of the phenomena of light. In keeping with our intent to discuss the historical development of gravitational wave production of electromagnetic radiation using modern notation, the equations describing electromagnetic radiation will be presented in a form that is completely covariant. The Maxwell equations are written in terms of tensor relations and will have the same form in Minkowski space and curved space-time. The physical properties of electromagnetic radiation, such as luminosity, will be developed in terms of Newman-Penrose scalars  \cite{Newman61,Teukolsky73}, in a form that is well suited for the comparison of electromagnetic and gravitational radiation \cite{Jones17}.

In order to write the field equations for electrodynamics in a suitable 
form for curved space-time two tensors are defined in terms of the 
electric and magnetic fields. The field strength tensor 
\cite{Ellis73,Senego98,Hogan09,Palenzuela10,Lehner09,Lehner12_85,
Lehner12_86,Lehner16} is defined as  \footnotemark \footnotetext{Great 
care is required with sign conventions in any covariant representation. 
This is particularly true in the case of Maxwell's equations and here we 
are following Palenzuela {\it et al.} \cite{Palenzuela10}, which is consistent
with our metric. It is prudent to check the signs by confirming that the 
covariant relations reduce correctly to the Maxwell equations in a 
Lorentz inertial frame.}

\begin{equation}
F^{\mu \nu }  = u^{\mu} E^{\nu}  - u^{\nu} E^{\mu} + e^{\mu \nu \alpha 
\beta} B_{\alpha} u_{ \beta},
\label{Faraday}
\end{equation}

\noindent and the dual to the field strength tensor as,

\begin{equation}
^*F^{\mu \nu }  = u^{\mu} B^{\nu}  - u^{\nu} B^{\mu}  - e^{\mu \nu 
\alpha \beta} E_{\alpha} u_{ \beta},
\label{dFaraday}
\end{equation}

\noindent where $e^{\mu \nu \alpha \beta}$ is the ``Levi-Civita 
pseudotensor of the space-time" and $u_{\nu}$ is the field frame 4-
velocity. The covariant expression for the field strength tensor 
\eqref{Faraday} and the dual \eqref{dFaraday} was originally developed 
by Ellis \cite{Ellis73}. While these expressions are perhaps not widely 
known, expanding \eqref{Faraday} in a Lorentz inertial frame produces 
the expected components for the field strength tensor. This covariant 
form of the field strength tensor has proven to be very useful in 
studies of the relation between gravity and electromagnetism 
\cite{Hogan09,Palenzuela10,Lehner09,Lehner12_85,Lehner12_86,Lehner16}. 
Conversely, the electric and magnetic fields are found from the 
contractions of the tensors with the 4-velocity,

\begin{equation}
E^{\mu} = F^{\mu \nu } u_{ \nu},~ ~ B^{\mu} = {^*}F^{\mu \nu } u_{ \nu}.
\label{EBfieldsF}
\end{equation}

\noindent Using the field strength tensor and its dual the Maxwell 
equations can be written in a  covariant form that is the same in both Minkowski space and curved 
space-time. The inhomogeneous Maxwell equations ({\it i.e.} Gauss's law and Ampere's) 
law are,

\begin{equation}
\frac{1}{\sqrt{-\left| g_{\mu \nu} \right|}} \partial_{\nu} \left( \sqrt{-\left| g_{\mu \nu} \right|} F^{\mu \nu } 
\right)=  4 \pi J^{\mu}.
\label{Maxwell1}
\end{equation}

\noindent where $ \left| g_{\mu \nu} \right|= det [g_{\mu \nu}]$ is the determinant of the metric. The homogeneous 
Gauss's law for magnetism and Faraday's law are \cite{Greiner96, 
Palenzuela10},

\begin{equation}
\partial_{\nu} \left( \sqrt{-\left| g_{\mu \nu} \right|} ~{^*}F^{\mu \nu } \right)  = 0.
\label{Maxwell2}
\end{equation}

\noindent The covariant form of the conservation law is,

\begin{equation}
\partial_{\mu} \left( \sqrt{-\left| g_{\mu \nu} \right|} J^\mu \right)= 0.
\label{Conservation}
\end{equation}

\noindent The field strength tensor can also be expressed in terms of 
the electromagnetic 4-vector potential,

\begin{equation}
F_{\mu \nu } = \partial_{\mu} A_{\nu} - \partial_{\nu} A_{\mu} ~.
\label{FourPotential}
\end{equation}

\noindent Maxwell came to the wave equation from the bottom up by recognizing that the displacement current term was missing from the traditional form of Ampere's law. In the modern notation the wave equation is a mathematical identity in the absence of source terms in the Maxwell equations \cite{Tsaga05}. 

The covariant form of the equations for the electromagnetic field appears naturally in the radiative expression for electrodynamics in the Newman-Penrose formalism \cite{Newman61,Teukolsky73} through the introduction of the Newman-Penrose electromagnetic scalar, to be discussed shortly. In order to provide a means of comparison between electromagnetic and gravitational radiation using the Newman-Penrose formalism we will require the Lagrangian density for the electromagnetic field in curved space-time. Including the electric source terms the Lagrangian density is,

\begin{equation}
 \mathcal{L}_
{em}  =  - \frac{1}{4}\left( {\partial _\nu  A_\mu   - \partial _\mu  A_\nu  } \right)\left( {\partial ^\nu  A^\mu   - 
\partial ^\mu  A^\nu  } \right) +J_\mu A^\mu.
\label{emLagrangian}
\end{equation}

\noindent The Lagrangian density can be simplified using the Lorenz gauge \cite{Greiner96}, $\partial_\mu A^\mu = 0$, and by restricting our attention to be source free ({\it i.e.} $J_\mu =0$) so that \eqref{emLagrangian} becomes,

\begin{equation}
 \mathcal{L}_{em}  =  - \frac{1}{2}\partial _\mu  A_\nu  \partial ^\mu
 A^\nu .
\label{emLagrangianLG}
\end{equation}

\noindent Since we are only considering the radiation regime for the 
field equations, we assume a plane wave solution for the 
electromagnetic field and a massless vector field can then be 
expressed in terms of a mode expansion \cite{Greiner96} as follows,

\begin{equation}
A_\mu  \left( {k,\lambda ,x} \right) = \epsilon_{\mu} 
^{(\lambda)} \phi ^{(\lambda)} \left( {k ,x} \right).
\label{ModeExpN}
\end{equation}

\noindent with $k$ being the momentum, $x$ being the 
space-time coordinates and $\epsilon ^{(\lambda)}$ being the 
polarization vectors with $\lambda = 0, 1, 2, 3$. For example, the two 
transverse modes are often labeled $\lambda =1 , 2$ with $\epsilon 
_\mu ^{(1)} = (0,1,0,0)$ and $\epsilon _\mu ^{(2)} =  (0,0,1,0)$. The 
$\lambda = 0, 3$ components are the time-like and longitudinal polarizations. The
four polarization vectors satisfy the orthogonality relationship
$\epsilon _\mu ^{(\lambda)} \epsilon ^{\mu ~(\lambda')} = 
\eta ^{\lambda \lambda'}$, with $\eta ^{\lambda \lambda'}$ being the
Minkowski metric.

One can define a complex scalar field using the $\lambda = 1, 2$ 
components of the real scalar fields {\it i.e.} 
$\phi ^{(1,2)} (k, x)$ as $\varphi =\frac{1}{\sqrt{2}}
(\phi ^{(1)} + i \phi ^{(2)})$. In terms of this complex scalar field 
the Lagrange density of \eqref{emLagrangianLG} can be written as
\begin{equation}
\label{emLagrangian2}
{\cal L}_{em} = - \partial_\mu \varphi \partial ^\mu \varphi ^* ~,
\end{equation}
which is the Lagrange density for a massless complex scalar field. Below we will use a massless,
complex scalar field as a stand-in for the massless photon. This substitution is justified 
by equation \eqref{emLagrangian2}

The field equations following from the Lagrange density in \eqref{emLagrangian2} are,   

\begin{equation}
\frac{1}{{\sqrt{-\left| g_{\mu \nu} \right|}}} \partial_{\mu} \sqrt{-\left| g_{\mu \nu} \right|} g^{\mu \nu} \partial_{\nu}\varphi = 0.
\label{eomvarphi}
\end{equation}

\noindent This expression for the field equations is the same for both curved space-time and for Minkowski space-time. If we restricted our attention to plane waves propagating in Minkowski space-time the solution to \eqref{eomvarphi} takes the form, 

\begin{equation}
\varphi = B e^{i  \text{k} u} + C,
\label{PlaneWave}
\end{equation}

\noindent where the $z$ axis is assumed to be along the direction of propagation, $B$ and $C$ are constants, and $k$ is the wave number. The solution \eqref{PlaneWave} is written in the standard light cone coordinate $u = z - t$ and $\partial_{z} - \partial_{t} = 2 \partial_{u}$ \cite{Jones16}.

The previous discussion provides an outline of Maxwell's electromagnetic radiation in modern covariant form. The principle goal of this review is to realize Faraday's expectation of a relationship between gravity and electricity. We will demonstrate this relationship for gravitational and electromagnetic radiation by comparing the gravitational radiation luminosity to the luminosity of the corresponding electromagnetic radiation produced from the vacuum by the gravitational radiation. We have found that the Newman-Penrose formalism is a good method for calculating the luminosity of both gravitational radiation and the corresponding electromagnetic radiation. The electromagnetic luminosity is presented here in terms of the Newman-Penrose formalism and the gravitational radiation luminosity will be presented in the following section.

The radiated electromagnetic power per unit solid angle is found from the projection of the electromagnetic field strength onto the elements of a null tetrad ($l_\mu, m^\mu, n^\mu , \bar m^\mu$) which gives us the electromagnetic Newman-Penrose scalar $\Phi _2$. In the Newman-Penrose formalism the power per unit solid angle of emission for electromagnetic radiation is written as \cite{Teukolsky73,Lehner09},

\begin{equation}
\frac{d E_{em} }{dt d\Omega}  = \mathop {lim}\limits_{r \to \infty } \frac{{r^2 }}{{4\pi }}\left| {\Phi _2 } \right|^2.
\label{emFlux0}
\end{equation}

\noindent The Newman-Penrose electromagnetic scalar in \eqref{emFlux0}  \cite{Newman61,Teukolsky73,Lehner09,Lehner12_86} is,

\begin{equation}
\Phi _2  = F_{\mu \nu } \bar m^\mu  n^\nu.
\label{Phi2}
\end{equation}

\noindent The null tetrad of the Newman-Penrose formalism in \eqref{Phi2} can be defined as \cite{Lehner12_85},

\begin{equation}
\begin{array}{*{20}c}
   {l^\mu   = \frac{1}{{\sqrt 2 }}\left( {1,0,0,1} \right),} & {n^\mu   = \frac{1}{{\sqrt 2 }}\left( {1,0,0, - 1} \right),}  \\
   {m^\mu   = \frac{1}{{\sqrt 2 }}\left( {0,1,i,0} \right),} & {\bar m^\mu   = \frac{1}{{\sqrt 2 }}\left( {0,1, - i,0} \right),}  \\
\end{array}
\label{null1}
\end{equation}

\noindent and

\begin{equation}
 l \cdot n =  - 1,~~m \cdot \bar m = 1 ,~~
l \cdot l = n \cdot n = m \cdot m = \bar m \cdot \bar m = 0.
\label{null2}
\end{equation}

\noindent The electromagnetic field strength tensor in \eqref{Phi2} is $F_{\mu \nu }  = \partial _\mu  A_\nu  {\kern 1pt}  - \partial _\nu  A_\mu$, where $A_\mu   = \epsilon ^{(\lambda)}  _\mu \phi ^{(\lambda )} \left( {t,z} \right)$, and the plane polarization vectors are $\epsilon ^{(1)} _\mu   = \left(0, 1, 0, 0 \right), \;\epsilon ^{(2)} _\mu   = \left(
0, 0, 1,0 \right)$ \cite{Jones17}. The electric and magnetic fields are determined by taking the derivatives of the scalar field \eqref{PlaneWave}: $\,\partial _t \varphi  =  - i  k B e^{i k \left( {z - t} \right)}$ and $\partial_z \varphi = i k B e^{i  k \left( {z - t} \right)}$. Collecting terms for the Newman-Penrose scalar of the ``out" state of \eqref{PlaneWave} \cite{Lehner12_85,Jones17},

\begin{equation}
\Phi _2  = F_{\mu \nu } \bar m^\mu  n^\nu =
\frac{1}{{\sqrt 2 }}e^{ - i\frac{\pi }{4}} \left( { \partial _z \varphi - \partial _t \varphi} \right) =
i e^{ - i\frac{\pi }{4}} {\sqrt 2 }  k B e^{i k \left( {z - t} \right)}.
\label{emNPscalar0}
\end{equation}

\noindent The square of the electromagnetic scalar is then,

\begin{equation}
\left| {\Phi _2 } \right|^2  = 2  k^2 B^2.
\label{emNPscalar}
\end{equation}

The square of the Newman-Penrose scalar in  \eqref{emNPscalar} is proportional to the electromagnetic flux, $F_{em} \sim  \left| {\Phi _2 } \right|^2 $. In Section \ref{gRad} on gravitational radiation and in Section \ref{production} on scalar field production we will show that by using the Newmam-Penrose scalars for the gravitational and the electromagnetic fields respectively, one can compare the fluxes of gravitational and electromagnetic radiation \cite{Jones17}.

\section{Gravitational radiation}  \label{gRad}

A wave like solution to the vacuum equations for general relativity exist similar to that of electromagnetism \cite{Schutz00}. This was recognized by Einstein soon after the development of general relativity and proposed even earlier by Poincar{\' e} \cite{Smoot16}. Initially there was doubt as to whether or not gravitational waves were physically real. Unlike the production of electromagnetic radiation there is no dipole source for gravitational radiation. This is because mass dipole production of radiation would violate conservation of 4-momentum \cite{Schutz00,Smoot16}. However, there are also quadrupole source terms which lead to a wave solution and does not violate any conservation principles \cite{Schutz00}.

Since our interest here is in the relation between gravity and electromagnetism, in the radiation regime, we will restrict our attention to the plane wave solution of general relativity. The metric of a gravitational plane wave traveling in the $+z$ direction and with the $+$ polarization
can be written as \cite{Schutz},

\begin{equation}
ds^2 = g_{\mu \nu} dx^{\mu} dx^{\nu} = -dt^2 + dz^2 + f^2 dx^2 + g^2 dy^2, 
\label{GWmetric}
\end{equation}

\noindent were we set $c=1$. This metric is oscillatory with $f = 1 + \varepsilon (u) $, $g = 1 - \varepsilon(u)$ with $\varepsilon(u) =h_+ e^{iku}$.  The coefficient $h_+$ is the gravitational wave strain amplitude and $k$ is the wave number. The coordinate variable in the metric is the standard light cone coordinate $u=z-t$. This metric only includes the ``plus" polarization. Similar to electromagnetic radiation there are two degrees of freedom corresponding to two polarization states for gravitational radiation. The two polarization states for gravitational radiation are ``plus" and ``cross" polarization. They differ by an angle of $\frac{\pi}{4}$ in contrast to a phase angle difference of $\frac{\pi}{2}$ for electromagnetism \cite{Schutz00, Schutz09}. Including the ``cross" polarization would not change our discussion.

\begin{figure}[H]
\centering
{\caption{The ``discovery paper" spectrogram of gravitational waves produced by a binary black hole in-spiral \cite{LIGO}.} \label{spectrogram}}
{\includegraphics[width=12cm]{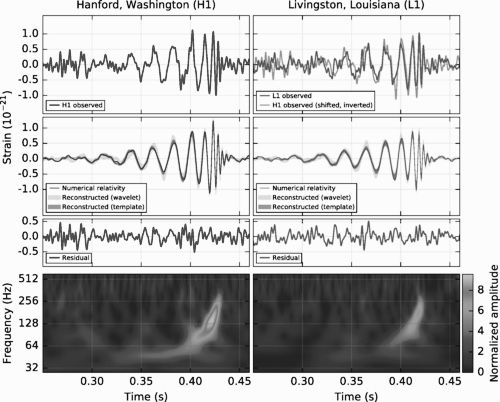}}
\end{figure}

The luminosity of the gravitational radiation can be calculated using the Newman-Penrose formalism \cite{Teukolsky73,Lehner09}. The relevant scalar for gravitational radiation is a projection of the Riemann tensor onto elements of a null tetrad. This projection is identified as the gravitational Newman-Penrose scalar $\Psi _4$. The power per unit of solid angle for the gravitational radiation is written in terms of the gravitational scalar as,

\begin{equation}
\frac{d E_{gw} }{dt d\Omega}  = \mathop {lim}\limits_{r \to \infty } \frac{{r^2 }}{{16\pi k^2 }}\left| {\Psi _4 } \right|^2 .
\label{GWluminsity}
\end{equation}

\noindent Substituting the metric for the gravitational wave \eqref{GWmetric} into the Riemann tensor for an outgoing gravitational plane wave in vacuum the gravitational Newman-Penrose scalar \cite{Teukolsky73,Jones17} is,

\begin{equation}
\Psi _4  =  - R_{\alpha \beta \gamma \delta } n^\alpha  \bar m^\beta  n^\gamma  \bar m^\delta   = f\partial_{u}^{2} f- g\partial_{u}^{2} g .
\label{Psi}
\end{equation}

\noindent The partial derivatives, $\partial_{u}$, are with respect to the light cone coordinate, $u$. For the plane wave metric, where $\varepsilon = h_+ e^{iku}$, the gravitational Newman-Penrose scalar and the square are given by,

\begin{equation}
 \Psi _4  =  - 2 h_+ k^2 e^{ik\left( {z - t} \right) } \rightarrow |\Psi _4 | ^2 = 4 h_+ ^2 k^4.
 \label{Psi4}
 \end{equation}

\noindent The Newman-Penrose scalars in \eqref{emNPscalar} and \eqref{Psi4} provide a convenient method for the comparison of the power of the gravitational radiation and the counterpart vacuum production of electromagnetic radiation which will follow. The flux of the gravitational wave can be calculated from \eqref{GWluminsity} and  \eqref{Psi} as \cite{Schutz,Schutz09,Gretarsson18},

\begin{equation}
F_{gw}= \frac{c^3}{16 \pi G} \left| \dot \varepsilon \right| =  \frac{c^3 h_+ ^2 \omega^2}{16 \pi G},
 \label{GWfluxh}
 \end{equation}
 
 \noindent which is a function of the strain amplitude $h$ and gravitational wave frequency $\omega = 2 \pi f$.
 
This review of the relationship between gravity and electromagnetism in the radiation regime might be considered of academic interest only except for the recent and remarkable discovery by the LIGO scientific collaboration of gravitational waves \cite{LIGO}. The spectrogram from the discovery papers is provided in Fig. \ref{spectrogram} for the binary in-spiral of two black holes. The detection of gravitational waves makes the discussion of the potential production of electromagnetic radiation by gravitational waves immediately relevant to current research in both the fundamental relation between gravity and electromagnetism as well as potential applications in astrophysics. If there were any lingering doubt about the certainty of the detection of gravitational waves the more recent detection of gravitational waves from a kilonova event \cite{ligo2} has laid these doubts to rest. The kilonova event was first identified through the detection of gravitational waves by the LIGO scientific collaboration.  The kilonova was immediately verified across the electromagnetic spectra through the coordination of an international collaboration of observatories based around the World and in space. The kilonova observations have not only eliminated any reasonable doubt of the existence of gravitational waves but also ushered in a new era of ``multi-messenger" astronomy and astrophysics.

\section{Electromagnetic fields in gravitational wave background}

\subsection{Gravitational waves and uniform magnetic field}

The first modern attempt to connect gravitational radiation and 
electromagnetic fields was work by Gertsenshtein \cite{Gertsenshtein60}
which considered the linearized Einstein field equations coupled to an 
electromagnetic plane wave.

\begin{equation}
\label{gert-1}
\Box {\tilde h}^{\mu \nu} = -\kappa t^{\mu \nu} ~,
\end{equation}

\noindent where $t^{\mu \nu} = \frac{1}{4 \pi} (F^{\mu \tau} F^\nu _\tau -
g^{\mu \nu} F^{\alpha \beta} F_{\beta \alpha})$ is the energy-momentum
tensor for the electromagnetic field, ${\tilde h}^{\mu \nu} = h^{\mu \nu} 
- \frac{1}{2} g^{\mu \nu} h$ is the trace reduced metric deviation 
of the metric tensor ({\it i.e.} $g^{\mu \nu} = \eta ^{\mu \nu} + h^{\mu 
\nu}$), and $\kappa = 16 \pi G$. Now the proposal in 
\cite{Gertsenshtein60} was to generate gravitational waves by sending 
electromagnetic waves through a constant magnetic field. This potential 
effect was compared to radio physics phenomenon of wave resonance. The 
idea being that despite the weak coupling of gravity one could 
nevertheless generate some significant amount of gravitational radiation 
by this method. 

Now if one takes the electromagnetic field to have a constant magnetic 
field part (whose field strength tensor we denote by $F^{(0) \mu \nu}$) 
and a plane wave part (whose field strength we denote by $F ^{\mu \nu}$),
and if we feed this into \eqref{gert-1}, dropping the squared terms in
$F^{(0) \mu \nu}$ and $F^{\mu \nu}$ and keeping only the cross terms 
we arrive at

\begin{equation}
\label{gert-2}
\Box {\tilde h}^{\mu \nu} = -\frac{\kappa}{2} \left( F^{(0) \mu \tau} 
F^{\nu} _\tau - \frac{1}{4} g^{\mu \nu} F^{(0) \alpha \beta} F_{\alpha \beta}    \right) ~.
\end{equation}

\noindent One now assumes that the electromagnetic plane wave field and
gravitational field propagate along the $z$ direction with wave number 
$k$ and have the form

\begin{equation}
\label{gert-3}
F^{\mu \nu} = b(z) \epsilon^{\mu \nu} e^{i(kz-\omega t)}  ~~~;~~~
{\tilde h}^{\mu \nu} = a(z) \zeta ^{\mu \nu} \sqrt{\frac{\kappa}{k^2}}
e^{i(kz - \omega t)} ~,
\end{equation}

\noindent where $\epsilon ^{\mu \nu} , \zeta ^{\mu \nu}$ are the 
electromagnetic and gravitational polarization tensors respectively. 
Using \eqref{gert-3} in \eqref{gert-2} and assuming slowly varying
amplitudes $a(z) , b(z)$ one obtains the following relationship 
between the amplitudes
\begin{equation}
\label{gert-4}
i \frac{da (z)}{dz} = \sqrt{\frac{\kappa}{16}} F^{(0) \mu \nu}
\epsilon_{\beta \nu} \zeta^{\beta} _\mu b(z) ~.
\end{equation}
Under the assumption that $b(x) \approx const.$ \eqref{gert-4} can be 
integrated to obtain $a(x)$ as
\begin{equation}
\label{gert-5}
\left| \frac{a (z)}{b(0)} \right|^2 = \frac{\kappa}{16 \pi^2} 
B_0 ^2 T^2 ~,
\end{equation}
where $B_0 \simeq | F^{(0) \mu \nu} |$ is the constant magnetic field 
strength, $T$ is the time that the electromagnetic wave traverses the
uniform magnetic field, and $b(0)$ is the initial amplitude of the
electromagnetic wave. If one takes the cosmological sized magnetic fields
($B_0 \simeq 10 ^{-5}$ G) and assumes cosmological times for the 
electromagnetic wave to travel through this constant magnetic field
($T \simeq 10^7$ years) one finds that the ratio of gravitational to 
electromagnetic amplitude is of the order $|a/b|^2 \simeq 10^{-17}$. One 
could increase this by having stronger magnetic fields and/or longer 
periods of travel. 

One of the most interesting features of the above mechanism is that the
gravitational wave frequency is the same as that of the electromagnetic 
wave. This gives the possibility of generating very high frequency 
gravitational waves compared to the ``natural'' sources of gravitational 
waves from the first series of direct detections-- merging black hole, 
merging neutron stars. These natural or astrophysical sources have 
frequencies in the 100s to 1000s of Hertz, whereas electromagnetic 
radiation has a much broader range of frequencies which have been
observed -- from 1000s of Hertz to Gigahertz and beyond.

In the original work by Gertsenshtein \cite{Gertsenshtein60} the focus 
was on generating gravitational waves from electromagnetic waves. In this
review our focus is the exact opposite -- we are interested in 
electromagnetic radiation generated from gravitational waves. This 
reversed possibility was pessimistically noted by Gertsenshtein with the
concluding comment ``From general relativity follows also the possibility
of the inverse conversion of gravitational waves into light waves, 
but this problem is hardly of interest.'' Nevertheless, several years 
after Gertsenshtein's paper, Lupanov \cite{lupanov67} did examine the 
inverse process of generating electromagnetic waves from gravitational
waves.  

We will follow the work in \cite{lupanov67} by examining the reverse 
process of electromagnetic radiation generated from gravitational waves. 
There are two  reasons for our focus on the reverse process: (i) 
electromagnetic radiation, even weak radiation, is easier
to detect, and (ii) we argue, beginning in the next subsection, that the
conversion of gravitational waves to electromagnetic radiation occurs 
even in vacuum.

Before concluding this subsection we mention that there is more recent work
in the spirit of Gertsenshtein's work \cite{Gertsenshtein60}
where the magnetic field is replaced by a Bose-Einstein condensate
\cite{fuentes}. In this case the interaction of the gravitational
wave with the Bose-Einstein condensate is conjectured to lead to 
the creation of phonons, just as in Gertsenshtein's work the 
interaction of the gravitational wave with the magnetic field
lead to the creations of photons. This creation of phonons
with a Bose-Einstein condensate has been put forward as a 
potential alternative mechanism to interferometers like LIGO
to detect gravitational waves.

\subsection{Feynman diagram approach to gravitational and electromagnetic
radiation}

\begin{figure}[!h]
\centering
\includegraphics[width=110mm]{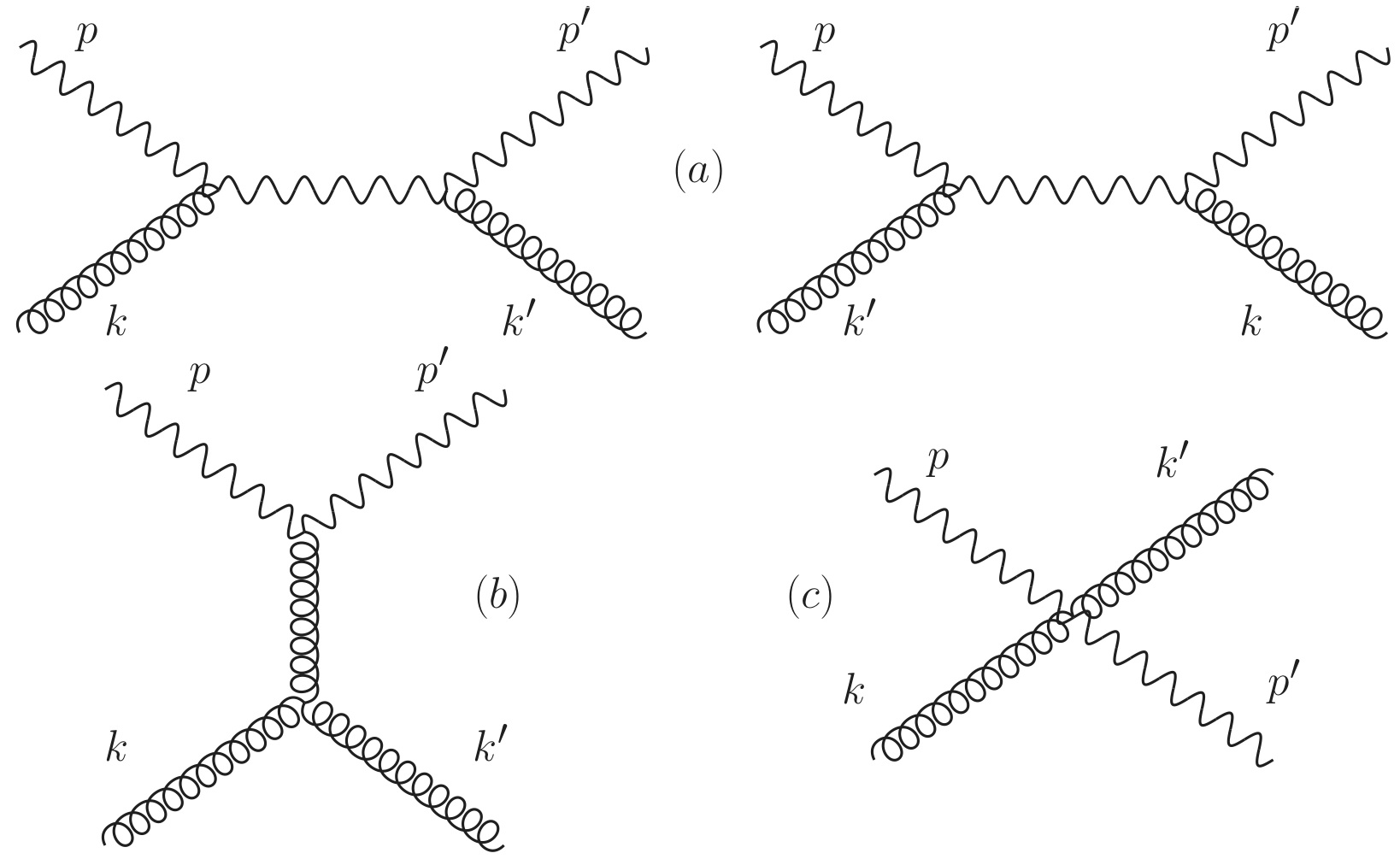}
\caption{Tree level Feynman diagrams of graviton-photon transitions 
\cite{Skobelev75}. Wavy lines represent photons and curly lines 
represent gravitons \label{Skobelev}}
\end{figure}

To check the assertion that electromagnetic radiation can be created in 
vacuum from gravitational radiation we turn to tree-level Feynman 
diagrams for graviton-photon scattering. The four basic diagrams for this 
process are given in Fig. \eqref{Skobelev} with curly lines 
representing gravitons and wavy lines represent photons. The original 
calculation was carried out by Skobelev \cite{Skobelev75} with more 
recent and more extensive calculations being found in references 
\cite{Bohr14}. The diagrams in Fig. \eqref{Skobelev} represent
$graviton + photon \to graviton + photon$ scattering. By rotating the 
diagrams one can get $graviton + graviton \to photon + photon$ or
$photon + photon \to graviton + graviton$ which can be viewed as 
creation of photons (gravitons) from gravitons (photons).

In \cite{Skobelev75} the process $graviton + photon \to graviton + 
photon$ is calculated first. After a long calculation the 
differential scattering cross section is found to be
\begin{equation}
\label{sko-1}
\frac{d \sigma}{d \cos \Theta} =\frac{\kappa ^2 \omega ^2}{64 \pi}
\left( \frac{1+ \cos ^8 (\Theta /2) }{\sin ^4 (\Theta / 2)} \right) ~,
\end{equation}
where $\kappa = 16 \pi G$ as previously, $\omega$ is the energy of 
the system, and $\Theta$ is the scattering angle.

Now the process of interest in this review is where gravitational waves 
create electromagnetic waves or photons. In the Feynman diagram language 
this means $graviton + graviton \to photon + photon$. This process 
can be obtained by rotating the diagrams in Fig. \ref{Skobelev}  by $90^0$ degrees. Upon 
doing this the differential cross section for $graviton + graviton \to 
photon + photon$ is found to be \cite{Skobelev75}
\begin{equation}
\label{sko-3}
\frac{d \sigma}{d \cos \Theta} =\frac{\kappa ^2 \omega ^2}{64 \pi}
(\cos ^8 (\Theta /2) + \sin ^8 (\Theta / 2) ) ~.
\end{equation}
Integrating \eqref{sko-3} to obtain the total cross section 
\begin{equation}
\label{sko-4}
\sigma =\frac{\kappa ^2 \omega ^2}{160 \pi} ~.
\end{equation}
If one takes the energy of the system to be the rest mass energy of 
the electron, $\omega = m_e c^2$, \footnote{This energy is much larger
than the energy implied by the frequencies of the observed gravitational 
wave signals \cite{LIGO}. For the energy associated with the frequencies
implies by the LIGO observations the cross section would be even smaller
} one finds \cite{Skobelev75} that  $\sigma \simeq 10^{-110} cm ^2$. 
This is a very small number and indicates that at the level of individual
photons and gravitons this is not a large effect. However, given the 
enormous energy of the observed gravitational wave signals, which implies a
large number of gravitons, we will argue that there are cases where the 
small cross section of \eqref{sko-4} may be compensated for by the 
large number of gravitons/strength of the gravitational wave. 

\subsection{Massive Scalar field in Gravitational Plane Wave Background}

The idea of particle creation from a time dependent space-time was 
first considered in a series of papers by Parker \cite{parker-1,parker-2,parker-3} which investigated particle production 
from the time dependent FRW cosmological space-time. The next major 
time-dependent space-time to be studied in terms of particle production 
was the gravitational plane wave space-time. The initial studies were 
carried out by Gibbons \cite{gibbons} and Deser \cite{deser}, who 
considered the  production of a massive scalar field in a pulsed 
gravitational wave space-time. Reference \cite{garriga} has a 
more extensive discussion of particle creation from a gravitational 
wave background, again in the context of a massive scalar field. 
The conclusion of all of these works was that massive scalar 
fields would not be created from such a plane wave  gravitational 
background. 

This conclusion, of no particle creation from a plane gravitational 
wave background, appears at odds with recent work \cite{Jones15,Jones16,Jones17,Jones18}. However these recent
works focus on the case of massless particles whereas references
\cite{gibbons,deser,garriga} focus on massive particles.
As noted by Gibbons one can already guess that the production of a 
massive field from a gravitational plane wave would be forbidden
by energy momentum conservation. A massless graviton can not 
decay/transform into massive particles since this would violate 
energy-momentum conservation. It is the same reason that forbids 
a photon from decaying into an electron-positron pair in free space.
(A photon can decay/transform into electron-positron pair in the
presence of a heavy nuclei which acts to conserve energy and momentum). 

General studies of when one type of massless field quanta can 
decay/transform into other massless field quanta can be found
in \cite{Modanese95} and \cite{Fiore96}. Using energy-momentum 
conservation these two works show that some number of massless
quanta can transform into some number of other massless field quanta
so long as the incoming and outgoing particles lie along the same 
direction. This is consistent with the Feynman diagram calculations
of reference \cite{Skobelev75} where the decay of gravitons to photons
({\it i.e.} $graviton + graviton \to photon + photon$) is possible so long as the
momenta of all particles lie along the same direction. This is also the
condition under which the creation of massless fields/field quanta 
occurs in references  \cite{Jones15,Jones16,Jones17,Jones18}.

In the rest of this section we review the calculations of \cite{gibbons,deser,garriga} which demonstrate the absence of
particle creation when the 
particles are massive, since this will provide a nice segue to
the case of massless particles. We will follow the notation of
reference \cite{garriga}.

Garriga and Verdaguer \cite{garriga} begin by considering the plane
wave metric of the form
\begin{equation}
\label{garr-1}
ds^2 = -dt^2 + dz^2 + g_{ab} (z, t) dx^a dx^b = 
-du dv + g_{ab} (u) dx^a dx^b ~,
\end{equation}
where in the last expression the metric has been transformed to light
front coordinates defined as $u=z-t$ and $v=z+t$ with $c=1$. The 
indices $a, b =1, 2$ and run over the $x, y$ directions, which are 
perpendicular to the $+z$ direction of travel of the gravitational wave.
For a gravitational wave traveling in the $-z$ direction one would take
the metric components to be functions of the light front coordinate
$v$ {\it i.e.} $g_{ab} (v)$.

Next a massive scalar field, $\varphi$, is placed in the metric given by 
\eqref{garr-1}. The equation for a massive scalar field in a curved
background is given by
\begin{equation}
\label{garr-2}
\left [ \frac{1}{{\sqrt { -\left| {g_{\mu \nu } } \right|} }}\left( 
{\partial _\mu  g^{\mu \nu } \sqrt { - \left| {g_{\mu \nu } } \right|} 
\partial _\nu  } \right) - m^2 \right] \varphi  = 0.
\end{equation}
Using  the metric \eqref{garr-1} in equation \eqref{garr-2}
and applying separation of variables, one finds solutions for the 
scalar field of the form
\begin{equation}
\label{garr-3}
\varphi (u, v, x^a)  = 
\frac{1}{(p_{-})^{1/2} (det g_{ab} (u))^{1/4} (2 \pi)^{3/2} }
\exp \left[i p_a x^a - i  p_{-} v -\frac{i}{4 p_{-}} 
\int _0 ^u (g_{ab} p^a p^b +m^2) du\right]~,
\end{equation}
with $p_{-}$ and $p_a$ being separation constants which physically
correspond to momenta connected with the coordinates,
$v$ and $x^a$ respectively. 

The next step is to calculate the Bogoliubov  coefficients \cite{davies} 
for this scalar field for a sandwich space-time {\it i.e.} one has a 
plane wave space-time for $u_1 < u < u_2 <0$, sandwiched between two 
Minkowski space-times for $u<u_1$ and $u >u_2$. The 
``in'' and ``out'' states for this sandwich space-time are \cite{garriga}
\begin{eqnarray}
\label{garr-4}
\varphi ^{in} (u, v, x^a)  &=& 
\frac{1}{(p_{-})^{1/2} (2 \pi)^{3/2} }
\exp \left[i p_a x^a - i  p_{-} v -\frac{i}{4 p_{-}} 
(p_a p^a +m^2) u + i \Delta \right]~,\\
\varphi ^{out} (u,v, x^a) &=&
\frac{1}{(k_{-})^{1/2} (2 \pi)^{3/2} }
\exp \left[i k_a x^a - i  k_{-} v -\frac{i}{4 k_{-}} 
(k_a k^a +m^2) u \right]~,
\label{garr-5}
\end{eqnarray}
where $\Delta$ is a constant phase. For the exact expression for this
phase as well as for the full details of the calculation and some 
subtleties in the definition of the coordinates we refer the reader to
\cite{garriga}. The light front momentum $p_-$ and $k_-$ are given by 
\begin{equation}
\label{garr-5a}
p_- = \frac{\omega_p - p_z}{2} ~~~;~~~ k_-=\frac{\omega_k - k_z}{2} ~,
\end{equation}
with $p_z , k_z$ being the three-momentum in the $z$ direction and 
$\omega_p = \sqrt{{\bf p}^2 +m^2}$ and $\omega_k = \sqrt{{\bf k}^2
+m^2}$ are the energies associated with the wave solutions. 

From \eqref{garr-4} and \eqref{garr-5} one can calculate the
Bogoliubov beta coefficients to be
\begin{equation}
\label{garr-6}
\beta = - \langle \varphi ^{in} | \varphi ^{out ~*}\rangle \propto 
\delta (p_- + k_-) ~,
\end{equation}
with the Dirac delta being a function of $p_- , k_-$ coming from 
integration over $dv$. Now if the scalar field is massive it is easy 
to see, using the expressions for $\omega _p \omega_k$ in equation 
\eqref{garr-5a} that
$p_- + k_- \ne 0$ so that $\beta =0$. However, if $m=0$ and the 3-
momentum are in the same direction ${\bf p} = p_z = {\bf k} = k_z$
then $p_- + k_- =0$ and $\beta \ne 0$ meaning that production of 
the scalar field from the gravitational wave occurs. This is 
consistent with the Feynman
diagram calculations of \cite{Skobelev75,Bohr14} as well as the
discussion in term of energy-momentum conservation of 
particle decay/production/scattering of massless fields 
\cite{Modanese95,Fiore96}. In the next section we investigate 
in more detail the possibility of producing massless fields/particles 
from a gravitational wave background.   

\section{Particle Production from a Gravitational Wave Background}

In this section we review some recent work \cite{Jones15,Jones16,Jones17,Jones18} on the production of massless fields from
gravitational wave backgrounds. We use a massless scalar field
to carry out the analysis, but our results also apply to the more
realistic case of a massless vector field from the results and
discussion around equations \eqref{emLagrangian},  \eqref{emLagrangianLG}, 
\eqref{ModeExpN}, and  \eqref{emLagrangian2}.
We also look at the response of an Unruh-Dewitt 
detector in a gravitational plane wave background which supports the
picture of gravitons decaying/transforming into photons. 
            
\subsection{Scalar field production} \label{production}

We now repeat some of the calculations of the previous section but for a 
massless scalar field. We will follow the work of \cite{Jones16}. For
the gravitational plane wave background we take the metric of 
\eqref{garr-1} to have the more specific form

\begin{equation}
\label{jones-1}
ds^2 = -dt^2 + dz^2 + f^2 (z,t) dx^2 + g^2 (z,t) dy^2 = 
-du dv +  f^2 (u) dx^2 + g^2 (u) dy^2  ~,
\end{equation}
where we have again transformed to light front coordinates, $u, v$ and 
taken $c=1$. The form of the metric in \eqref{jones-1} assumes the
plus-polarization for the gravitational plane wave, which we take 
without loss of generality. We further assume that the ansatz functions
have a oscillatory behavior of the form $f(u) = 1 + h_+ e^{iku}$ and
$g(u) =1 - h_+ e^{iku}$, where $h_+$ is the dimensionless amplitude
of the plus polarization and $k$ is the gravitational wave 
number. With $m=0$ the field equation for $\varphi$ from 
\eqref{garr-2} becomes

\begin{equation}
\label{jones-2}
\frac{1}{{\sqrt { -\left| {g_{\mu \nu } } \right|} }}\left( 
{\partial _\mu  g^{\mu \nu } \sqrt { - \left| {g_{\mu \nu } } \right|} 
\partial _\nu  } \right)  \varphi  = 0.
\end{equation}

Using the plane wave metric from \eqref{jones-1} along with the oscillatory form of the ansatz functions $f(u), g(u)$ equation \eqref{jones-2}  becomes

\begin{equation}
 \left( {4F (ku) \partial _u \partial _v  - 4ikG (ku) \,\partial _v  + 
 H(ku) \left( {\partial _x^2  + \partial _y^2 } \right)} \right)\varphi  
 = 0,
\label{jones-3}
\end{equation}

\noindent where the functions $F(ku), G(ku), H(ku)$, are given by

\begin{equation}
\begin{array}{l}
 \quad F\left( {ku} \right)  \equiv  \left( 1 - h_ + ^2 e^{2iku}  \right) 
 ^2, \\ 
 \quad G\left( {ku} \right)  \equiv  h_ + ^2 e^{2iku} \left( 1  -
 h_ + ^2 e^{2iku}  \right), \\ 
 \quad H\left( {ku} \right)  \equiv  \left( {1 + h_ + ^2 e^{2iku} } 
 \right). \\ 
 \end{array}
\label{jones-4}
\end{equation}

 \noindent In arriving at \eqref{jones-3} we have assumed that the
 behavior of $\varphi$ in the perpendicular $x, y$ directions are the 
 same so that
 $\partial _x \varphi = \partial _y \varphi$ and 
 $\partial _x ^2 \varphi = \partial _y ^2 \varphi$.

To solve \eqref{jones-3} we employ separation of variables as
$\varphi (u, v, x, y) = U(u) V(v) X(x) Y(y)$. The ansatz functions
in the $v, x, y$ directions are plane waves of the form

\begin{equation}
 X (x) = e^{ik_{xy} x} ~~~;~~~ Y(y) = e^{ik_{xy} y} ~~~;~~~~ 
 V(v) =  e^{ik_v v} ~,
\label{jones-5}
\end{equation}

\noindent where we have enforced the equality of the $x$ and $y$ 
directions by taking a common wave number, $k_{xy}$. With this 
set up and the solutions from \eqref{jones-5} the solution for $U(u)$
is \cite{Jones16}
\begin{equation}
U = B e^{\frac{\lambda }{k}} e^{  \frac{- \lambda }{{k\left( {1 - h_ + 
^2 e^{2iku} } \right)}}} \left( {1 - h_ + ^2 e^{2iku} } 
\right)^{\frac{1}{2}\left( {\frac{\lambda }{k} - 1} \right)} e^{ - 
i\lambda u} + C ~,
 \label{jones-6}
\end{equation} 
\noindent with $B, C$ being integration constants and 
$\lambda = \frac{k_{xy}^2}{2 k_v}$. Putting equations
\eqref{jones-5} \eqref{jones-6} together, and taking $C=-B$ the 
scalar field in the plane wave background becomes
\begin{equation}
\varphi (u, v, x, y)   = B e^{\frac{\lambda }{k}} e^{ - \frac{\lambda }
{{k\left( {1 - 
h_ + ^2 e^{2iku} } \right)}}} \left( {1 - h_ + ^2 e^{2iku} } 
\right)^{\frac{1}{2}\left( {\frac{\lambda }{k} - 1} \right)} e^{ - 
i\lambda u} e^{ik_v v} e^{ik_{xy} x} e^{ik_{xy} y} - B.
\label{jones-7}
\end{equation}
In the limit $h_+ \to 0$ ({\it i.e.} the gravitational background is  
turned off) the scalar field in \eqref{jones-7} becomes
\begin{equation}
\varphi _0  (t,x,y,z) = 
B e^{ - i\lambda u} e^{ik_v v} e^{ik_{xy} x} e^{ik_{xy} y} 
-B \rightarrow
 B e^{i\left( {k_v  + \lambda } \right)t} e^{i\left( {k_v  - \lambda } 
 \right)z} e^{ik_{xy} x} e^{ik_{xy} y} -B ~,
 \label{jones-8}
\end{equation}
\noindent where in the last step we have converted back to the original
$t,x,y,z$ coordinates. One can see that $k_v + \lambda$ plays the role
of the wave energy and $k_v - \lambda$ momentum in the $z$ direction. 
The result in \eqref{jones-8} is expected, since if the gravitational
wave background is turned off one should recover a plane wave traveling
in free space, which is what the solution in \eqref{jones-8} represents. 

Taking the limit where all the wave numbers/momenta go to zero 
({\it i.e.} $k_v, \lambda, k_{xy} \to 0$) in equation \eqref{jones-7}
one would expect the scalar field to vanish. However on taking this
limit one finds instead that 
\begin{equation}
\varphi (u, v, x, y)   = B \left[\left( {1 - h_ + ^2 e^{2iku} } 
\right)^{-\frac{1}{2}} - 1 \right] \approx \frac{B}{2} h_ + ^2 e^{2iku}
+ \frac{3B}{8} h_ + ^4 e^{4iku} ~.
\label{jones-9}
\end{equation}
The result in \eqref{jones-9} shows that even when one tries to take
the wave to its vacuum state, namely $k_v , \lambda, k_{xy} \to 0$,
there is a non-vanishing and non-trivial scalar field. This 
non-vanishing scalar field is the field/field quanta created by the 
gravitational wave background. Note that if one takes $h_+ \to 0$
in \eqref{jones-9} that one does get the expected value for the scalar
field $\varphi \to 0$ \footnote{Setting the constant $C=-B$ in
\eqref{jones-6} is done to get $\varphi \to 0$ in this limit. If one takes 
$C=0$ the $h_+ \to 0$ limit of \eqref{jones-9} would be
$\varphi \to B$ which is also a vacuum solution to the wave equation
for the massless $\varphi$, but having $\varphi \to 0$ is more ``natural".}
The four-current associated with $\varphi$ is given by the standard
expression $j_\mu = -i (\varphi \partial _\mu \varphi ^* - \varphi^* 
\partial_\mu \varphi)$. Inserting this solution from \eqref{jones-7} in the expression for the four-current and time averaging gives \cite{Jones16}
\begin{equation}
\langle j_\mu \rangle = -2 B^2 \lambda - B^2 h_+ ^4 \left(
\frac{9}{2} \frac{\lambda ^3}{k^2} -\frac{12 \lambda ^2}{k}
+ \frac{13}{2} \lambda - k \right) ~.
\label{jones-10}
\end{equation}
The constant $B$ is determined by choosing a normalization condition or
convention. Following references \cite{stahl} and \cite{Jones16} 
we pick the normalization condition $B = \frac{1}{\sqrt{2 k V}}$. 
Other possible normalization conditions for $B$ are 
discussed in \cite{halzen}. With this normalization the vacuum 
scalar field from \eqref{jones-9} reads
\begin{equation}
\varphi (u, v, x, y)   =    \frac{1}{\sqrt{2 k V}}
\left[\left( {1 - h_ + ^2 e^{2iku} } 
\right)^{-\frac{1}{2}} - 1 \right] \approx 
\frac{1}{2 \sqrt{2 k V}} h_ + ^2 e^{2iku}
\left( 1 + \frac{3}{4} h_ + ^2 e^{2iku} \right) ~.
\label{jones-11}
\end{equation}
Time averaging this vacuum current from \eqref{jones-10} gives
\begin{equation}
\langle j_\mu \rangle =   \frac{\rm{sign} (k) h_+ ^4 }{2 V}~.
\label{jones-12}
\end{equation}

\noindent In \eqref{jones-11} we are using a normalization that assumes the 
scalar field is in a box of volume $V$. In the previous section we took  
$B=\frac{1}{(2 \pi) ^{3/2}}$ -- see \eqref{garr-4} \eqref{garr-5}.

Equation \eqref{jones-10} gives the effect, in terms of the 
four-current, of passing a massless scalar field through a 
gravitational wave. On setting all the energy-momentum of the
scalar field to zero one finds, from equations \eqref{jones-11} and 
\eqref{jones-12}, that the scalar field and scalar field current
do not vanish. This represents the production 
of scalar field/scalar field quanta from the gravitational wave
background. 

Following \cite{Jones17} the ratio of the 
produced electromagnetic radiation \eqref{emFlux0} 
to the gravitational \eqref{GWluminsity} radiation can be written 
down in terms of electromagnetic and gravitational Newman-Penrose
scalars from  \eqref{Phi2} and \eqref{Psi4},  

\begin{equation}
\frac{{dE_{em} }}{{dE_{gm} }} = \frac{{\left( {\frac{1}{{4\pi }}\left| {\Phi _2 } \right|^2 } \right)}}{{\left( {\frac{1}{{16\pi k^2 }}\left| {\Psi_4 } \right|^2 } \right)}} = \frac{F_{em}}{F_{gw}}.
\label{emgwRatioA}
\end{equation}

\noindent Switching to a normalization where $B=1$ in \eqref{jones-9} 
the amplitude of the leading term of the scalar field is $\frac{h_+^2}{2}$.
Using this in the expression for $|\Phi_2|^2$  calculated 
in \eqref{emNPscalar}  we get $|\Phi_2|^2 =  2 k^2 h_+^4$. 
Next from \eqref{Psi4} we recall that 
$|\Psi _4 |^2 = 4 h_+^2 k^4$ for a gravitational plane we. Using all this
in \eqref{emgwRatioA} yields a relationship between the electromagnetic wave 
flux and gravitational wave flux

\begin{equation}
F_{em}  = 2 h_+^2 F_{gw}.
\label{emgwRatioB}
\end{equation}

\noindent Since the gravitational radiation is proportional to $h_+^2$ the electromagnetic production will be proportional to $h_+^4$. The result in equation \eqref{emgwRatioB} is consistent with the result in equation
\eqref{jones-12}.

\subsection{Unruh-Dewitt detector approach}

Another approach to study the connection between gravitational and 
electromagnetic radiation is through the use of an Unruh-DeWitt detector
\cite{unruh-det,dewitt-det}. An Unruh-DeWitt detector is a two-state, 
quantum system which is placed in a given space-time background. If the
Unruh-DeWitt detector is excited from the low energy state to the 
high energy state, this is taken to indicate that the given space-time
has produced field quanta in order to excite this transition. 
Two common examples of the use of an Unruh-DeWitt detector are placing 
it in the Schwarzschild space-time \cite{davies,hawking} of a black hole or 
placing it the Rindler space-time of an accelerating observer \cite{unruh}. 
In the first case the Unruh-DeWitt detector will detect the photons from 
Hawking Radiation and in the second case the Unruh-DeWitt detector will 
detect the photons from Unruh Radiation.

In this subsection we will summarize the work of reference \cite{Jones15}
which calculates the response of an Unruh-DeWitt detector interacting with 
a plane gravitational wave. The expression for the spectrum, $S(E)$, 
of an Unruh-DeWitt detector is given by 
\begin{equation}
\label{UD-1}
S(E) = n_{general} - n_{inertial} = 2 \pi \rho (E) F(E) ~.
\end{equation}
In equation \eqref{UD-1} $n_{general}$ and $n_{inertial}$ are the
photon density of a general space-time and inertial space-time
respectively. The difference between these two is a measure of the
photons created due to the general space-time. The terms $\rho (E)$
and $F(E)$ are, respectively, the density of states and response 
function both as a function of energy \cite{davies,letaw,akhmedov,wilburn,rad}. 

The detector response function is given by
\begin{equation}
\label{UD-2}
F(E) = \int _{-\infty} ^{+\infty} e^{-i \Delta \tau \Delta E}
(G^+ _{general} (\Delta \tau) - G^+ _{inertial} (\Delta \tau))
d (\Delta \tau ) ~,
\end{equation}
where we recall that $\hbar =1$ and $c=1$ in the above
formulas. $\Delta E = E_{up} - E_{down}$ is the energy difference 
between the two states of the Unruh-DeWitt detector. For simplicity 
we assume $E_{down} = 0$ so that $\Delta E \to E_{up} \to E$ and 
thus the response function is written as just a function of $E$. 
The terms $G^+ _{general} (\Delta \tau)$ and  $G^+ _{inertial} 
(\Delta \tau))$ are the  Wightman functions \cite{davies,Jones15} for the detector path in a general space-time and 
the detector path in the inertial space-time. The
Wightman function depends on the proper time difference 
$\Delta \tau$ 
for the path through the given space-time. The space-time path for 
the inertial  detector is $x^\mu (\Delta \tau ) = (\Delta \tau , 0, 
0, 0)$. The Wightman function for this inertial detector is
\begin{equation}
\label{UD-3}
G_{inertial} ^+ = \frac{1}{4 \pi ^2 x^\mu x_\mu} =\frac{1}{4 \pi ^2 
\Delta \tau ^2} ~.
\end{equation}
For a gravitational wave traveling in the $+z$ direction and having
$+$ polarization the space-time path is given by $x^\mu (\Delta 
\tau ) = ( \gamma \Delta \tau , \Delta x, 0, 0)$, where $\Delta x$ 
is the spatial displacement of the detector due to the gravitational 
wave and $\gamma ^{-2} = 1 - \Delta {\dot x} ^2$. Without loss of 
generality the detector is taken to be aligned along the $x$ direction. 
The Wightman function for the gravitational wave is 
\begin{equation}
\label{UD-4}
G_{wave} ^+ = \frac{1}{4 \pi ^2 x^\mu x_\mu} =\frac{1}{4 \pi ^2 
(\gamma ^2 \Delta \tau ^2 - \Delta x^2)} ~.
\end{equation}
Using these two Wightman functions from \eqref{UD-3} \eqref{UD-4} in 
\eqref{UD-2} we find
\begin{equation}
\label{UD-5}
F(E) = \frac{1}{2\pi^2 }\int _0 ^{+\infty} \cos (E \Delta \tau)
\left( \frac{1}{ (\gamma ^2 \Delta \tau ^2 - \Delta x^2)}- 
\frac{1}{\Delta \tau ^2} \right)
d (\Delta \tau ) ~.
\end{equation}
To evaluate \eqref{UD-5} we need to give $\Delta x$ as a function of 
$\Delta \tau$. This is done using the ${\dot x} = (1 + \frac{1}{2}h)$ 
\cite{EFTaylor} which is the expression for the trajectory along a null 
geodesic, to first order, for a gravitational wave background 
characterized by the amplitude $h (\Delta \tau, \theta, \psi) = h_0 
\sin^2 (\theta) \sin (2 \psi ) \sin (\omega \Delta \tau )$. The angles
$\theta$ and $\psi$ give the orientation of the axis of the detector 
with respect to the incoming gravitational wave \cite{hendry}. The 
separation $\Delta x$ between the particle undergoing geodesic 
motion in the gravitational wave background characterized by 
$h (\Delta \tau, \theta, \psi)$ and an inertial observer is then
given by $\Delta x = (1 + \frac{1}{2}h) \Delta \tau - \Delta \tau =
\frac{1}{2}h (\Delta \tau, \theta, \psi) \Delta \tau$. Using these
results in \eqref{UD-5} and integrating over $\Delta \tau$ as well as 
integrating over the orientation direction $\theta$ and $\psi$ gives
the detector response function as \cite{Jones15}
\begin{equation}
\label{UD-6}
F(E) = \frac{3 \pi}{256} h_0 ^2 (2 \omega  - E)   \Theta (2 \omega - E),
\end{equation}
where $\Theta$ is the Heaviside step function. Thus $F(E) = 0 $ when
$E > 2 \omega$ which is a similar type of cut-off to that in 
muon decay \cite{halzen,griffiths}. This suggests a picture of
gravitons ``decaying" into photons -- $graviton + graviton \to photon + 
photon$ or $graviton \to graviton + photon + photon$.

Using the response function from \eqref{UD-6} and a density of
states $\rho (E) = \frac{E^2}{2 \pi^2}$ \cite{letaw} the spectrum can 
be found from \eqref{UD-1} as
\begin{equation}
\label{UD-7}
S(E) = \frac{3 }{256 \pi \hbar ^3 c^3} E^2 h_0 ^2 
(2 \hbar \omega  - E)  \Theta (2 \hbar \omega - E) ~.
\end{equation}
We have restored factors of $\hbar$ and $c$ temporarily. The
functional form of the spectrum from \eqref{UD-7} is that of a
$Beta (3,2)$ distribution which is reminiscent of particle decays. This
again supports the picture of gravitons decaying into photons. 

The analysis of the present subsection is different from the proceeding
subsection in that here we place an Unruh-DeWitt detector
in the presence of a gravitational plane wave background, whereas in
the previous subsection we focused on the response of the vacuum to 
a gravitational wave. The Unruh-DeWitt calculation is closer in spirit to 
the work of Gertsenshtein \cite{Gertsenshtein60} where the gravitational 
wave interacts with a magnetic field. In both these cases there is some 
physical object --  the Unruh-DeWitt detector or a magnetic field -- which 
interacts with the gravitational wave. In the previous subsection the 
gravitational wave interacts with the quantum vacuum. Nevertheless all of 
these calculations indicate that
a gravitational wave can create electromagnetic radiation, or in
particle language that gravitons can transform/decay into photons. The work in \cite{Calmet16}
also looks into this possibility of gravitons decaying/transforming in to  photons and thus
weakening the gravitational wave.

\section{Possible Observational Consequences/Signatures}

While gravitational waves have only been directly detected very recently, electromagnetic 
radiation has been observed for all of human existence. If gravitational waves produce 
counterpart electromagnetic radiation as is outlined above, it is natural to ask what the 
possible observable consequence of this would be. In this section we address two possible observational
consequence: (i) the attenuation/decay of the gravitational wave due to production electromagnetic radiation;
(ii) the direction detection of the electromagnetic radiation produced by the gravitational wave. 

\subsection{Decay/attenuation of the gravitational wave}

If electromagnetic waves are produced from a gravitational wave, as suggested above, this
should weaken and attenuate the gravitational wave since this electromagnetic radiation must be
created at the expense of the gravitational wave \cite{Jones15,Jones16}. This is similar 
to how a black hole is conjectured to lose mass as a result of Hawking radiation -- the 
Hawking radiation comes at the expense of the mass of the black hole. 

One can connect particle/field production rate, $\Gamma$, with a current, $j_\mu$, as in equation
\eqref{jones-12} via the relationship \cite{stahl,nikolic,frob}
\begin{equation}
    \label{decay-1}
    \frac{\Gamma}{V} \Delta T \approx | j_\mu | ~,
\end{equation}
where $\Delta T$ is some characteristic time for the system and $V$ is the volume. Using $|j_\mu| = \frac{h_+ ^4}{2 V}$
from \eqref{jones-12} and taking $\Delta T \approx \frac{1}{\omega}$ (where $\omega$ is the frequency of the gravitational
wave) as the characteristic time we arrive at
\begin{equation}
    \label{decay-2}
    \Gamma \approx \frac{\omega h_+ ^4}{2}~.
\end{equation}
If we denote the number of gravitons in the volume $V$ as, $N_g$ one can write out a rate of change of $N_g$ as
\begin{equation}
    \label{decay-3}
    \frac{d N_g}{dt} = - \Gamma N_g \to   c \frac{d N_g}{dz} = - \Gamma N_g ~.
\end{equation}
In the last step we have replaced $dt$ by $dz/c$ since we want the decay as a function of distance rather than 
time. Taking the number of gravitons to be proportional to the amplitude squared \footnote{This is similar to QED where
the number of photons is proportional to the square of the vector potential -- $N_\gamma \propto A_\mu A^\mu$}
({\it i.e.} $N_g \propto h_+ ^2$) and using the expression of $\Gamma$ from \eqref{decay-2} we arrive at an
equation for how the amplitude, $h_+$, varies with distance, $z$,
\begin{equation}
    \label{decay-4}
c \frac{d {h_+ ^2}}{dz} = - \frac{\omega h_+ ^4}{2} (h_+ ^2) \to \frac{dh_+}{dz} = - \frac{k h_+ ^5}{4}~.
\end{equation}
One can solve \eqref{decay-4} for $h_+ (z)$ and find
\begin{equation}
    \label{decay-5}
h_+ (z) = (k z + K_0) ^{-1/4}~,
\end{equation}
where $K_0 = (h_+ ^{(0)} ) ^{-4}$ and $h_+ ^{(0)}$ is the reference amplitude at $z=0$. From equation 
\eqref{decay-5} one sees that the fall off of $h_+$ as a function of distance, $z$, is very slow. This is expected,
since this slow fall off tells us that the transformation of gravitational radiation (gravitons) into electromagnetic
radiation (photons) is a very weak process. If a gravitational wave background did not produce electromagnetic radiation
then $h_+ (z)$ should remain constant (recall that in this plane wave approximation we do not take into account
the $\frac{1}{r}$ fall of a real three dimensional wave).  

To get an idea of how weak the effect is we can calculate the ``half-distance", $\Lambda$,  which we define 
as the distance for the amplitude of the plane wave to fall to half of its initial value, $h_+ ^{(0)}$.  
Taking $\omega \approx 300$ Hz, the approximate frequency of the signal from the first LIGO detection
\cite{LIGO}, gives $k = \frac{\omega}{c} = 10 ^{-6}$ m. Setting $h_+ (\Lambda) = \frac{1}{2} h_+ ^{(0)}$ 
gives the ``half-distance" as
\begin{equation}
    \label{decay-6}
\Lambda = \frac{15}{k (h_+ ^{(0)} )^4} = \frac{1.5 \times 10^7}{ (h_+ ^{(0)} )^4} \rm{m}~,
\end{equation}
If one sets the ``half-distance", $\Lambda$,  equal to the size of 
the observable Universe -- $\Lambda = 10^{27}$ m -- then 
equation \eqref{decay-6} gives an amplitude of $h_+ ^{(0)} 
\approx 10^{-5}$, which is a very large amplitude. Equation
\eqref{decay-6} implies that as $h_+ ^{(0)}$ gets larger the 
half-distance, $\Lambda$, gets smaller. Taking  $h_+ ^{(0)} 
\approx 10^{-3}$ would give $\Lambda \approx 10 ^{19}$ m, which 
is 100 times smaller than the size of the Milky Way. We also want to stress again that
the above estimates based on equation \eqref{decay-6} are for planes waves and do not take into account the
$\frac{1}{r}$ fall off of a more realistic three dimensional wave, but regardless they
indicate that the decay/attenuation of the gravitational wave due to vacuum production of
electromagnetic radiation is a small effect, except perhaps close to the source where 
one might have amplitudes like $h_+ ^{(0)} \approx 10^{-3}$ or larger.  

\subsection{Detection of electromagnetic radiation produced by gravitational waves}

Next we look at the possibility of directly detecting the electromagnetic radiation that is
produced from the vacuum by the gravitational wave. Looking at \eqref{jones-9} one can 
see that the counterpart electromagnetic radiation production would have twice the
frequency of the gravitational wave. From \eqref{jones-9} one can see there are also components that
are at four times the frequency of the gravitational wave, but they are down by an extra factor of
$h_+ ^2$ compared to the component at twice the frequency. 

The first problem that occurs in potentially detecting the counterpart electromagnetic radiation is that
it will have a very low frequency (VLF) and thus a very large wavelength. For example the discovery paper 
\cite{LIGO} reported frequencies for the gravitational wave on the order of $100~\rm{Hz}$. Even doubling this,
the electromagnetic radiation would have a frequency and wavelength of  $200~\rm{Hz}$ and $1.5 \times 10 ^6 \rm{m}$
respectively. 

\begin{figure}[H]
\centering
\includegraphics[width=130mm]{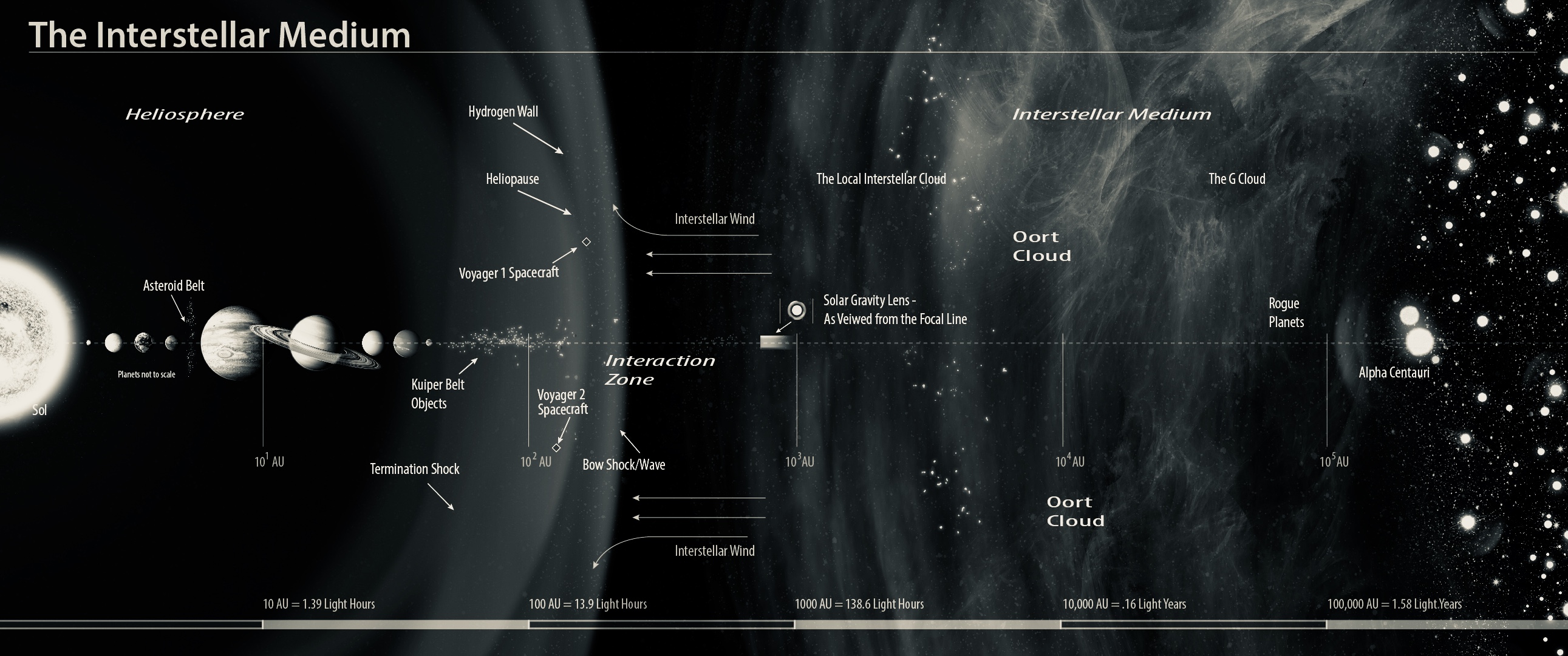}
\begin{tabular}{|c|c|}
\hline  ~~ Region ~~  & \  ~~ Observable frequency range ~~       
\\   
\hline  On Earth & \ $> 10 $ MHz          
\\  
\hline  Interplanetary space  (near Earth's orbit) & \ $>$ $20~\rm{kHz} - 30~\rm{kHz} $  
\\
\hline  Interstellar space (outside the heliosphere) & $ > 2 $ kHz 
\\   
\hline
\end{tabular}
\caption{Log scale illustration of the regions of space within our solar system and galaxy \cite{UCSB} \label{MediumFig} and the associated plasma frequency cutoff in each region \cite{Jones17,Lacki10}.}
\label{CutoffsFig}
\end{figure}

A second problem with detecting the VLF counterpart electromagnetic radiation is that there are various cutoff
frequencies due to the plasma in space. In the illustration and table below we give the plasma cutoffs for a detector
located in one of three locations: on the Earth, in space but near Earth's orbit, and finally in interstellar space
as shown in Fig. \eqref{CutoffsFig}. 

For a detector on Earth one can detect electromagnetic radiation with a frequency of $10~\rm{MHz}$ or larger. 
Assuming this electromagnetic wave came from production by a gravitational wave, this would require a
gravitational wave frequency of $5~\rm{MHz}$. Since non-exotic gravitational waves sources are expected to have 
frequencies that are orders of magnitude lower than this, this rules out an Earth based detector for such VLF electromagnetic 
radiation. 

For a detector at the outer edge of the Solar System, near interstellar space, one has a plasma cutoff of 
$2~\rm{kHz}$ which would require a gravitational wave frequency of $1~\rm{kHz}$. The fundamental ({\it i.e.} 
f-modes) of neutron star quakes have frequencies in the range $1 - 3~\rm{kHz}$ \cite{Kokkotas97} and thus upon 
doubling this frequency could produce counterpart VLF electromagnetic radiation above the $2~\rm{kHz}$ cutoff. 
In fact 
the Voyager probes did detect such VLF electromagnetic radiation \cite{Kurth84} in the range of $2-3 ~\rm{kHz}$
showing that detection of such VLF electromagnetic radiation is possible \footnote{The source of the Voyager 
detection of this VLF electromagnetic radiation was a mystery for some time, but the source of this
VLF radiation is now thought to be due to the interaction of the solar wind with ions in the outer heliosphere 
during times of intense solar activity \cite{Kurth03, Webber09}.} Thus
one could detect counterpart VLF electromagnetic radiation from
the $f$-modes of neutron star quakes if the neutron star were
close enough. However, it requires sending a probes to the edge of
the Solar System or beyond. 

Given that sending probes to the edge of the Solar System is costly both in terms of time and money 
one could ask if there are gravitational wave sources which would give rise to VLF electromagnetic 
radiation, which could be detected near Earth's orbit. From Fig. \eqref{CutoffsFig}  one can see that for
detection one needs the electromagnetic radiation to have a frequency greater than $20 - 30$ kHz. This implies that
the gravitational wave vacuum producing the electromagnetic radiation would need to have a frequency in the range
$10-15$ kHz. Theoretical models show that of gravitational wave of this kHz frequency range  could be 
produced from neutron star oscillations \cite{Kokkotas97,Kokkotas01}. There are different types of 
neutron star oscillation modes. Three
of the most common are: (i) {\it p}-modes or ``pressure modes" \cite{Kokkotas97} with a frequency 
range $5 - 9~ \rm{kHz}$; (ii) {\it f}-modes or  ``fundamental modes" \cite{Kokkotas97} with a
frequency range of $1 - 3~\rm{kHz}$; (iii) {\it w}-modes or ``space-time modes" \cite{Anderson96} with 
a frequency range of $8 - 16~\rm{kHz}$ or greater. From this list of oscillation modes 
the {\it w}-modes have the most promising frequency range in terms of detection of the 
counterpart VLF electromagnetic radiation. 

We now want to give a rough estimate of the strength of the electromagnetic flux produced by 
gravitational waves coming from a  {\it w}-mode oscillation of a neutron star quake. First from
\cite{abadie} the gravitational wave amplitude at Earth for {\it f}-mode generated gravitational waves 
from a neutron star that is $1$ kpc  or $3\times 10^{19}$ m distant from Earth would be of order
$h_+ \sim 10^{-23}$. The associated {\it w}-modes gravitational wave amplitude is expected to be
down from this by at least one order of magnitude $h_+ \sim 10^{-24}$. Using this {\it w}-mode 
amplitude and the $1/r$ fall off relation 
\begin{equation}
\label{detect-1}
    h_+ = 10^{-24} \frac{1 ~ {\rm kpc}}{r}
\end{equation}
we can determine the amplitude at some point close to the source. For this we take 
$r^{(0)}=3 \times 10^4$ m \cite{Jones17} -- this is far enough from the neutron star that the plane wave 
approximation we have used throughout should apply, at least roughly. Using \eqref{detect-1} and 
$r^{(0)}=3 \times 10^4$ m we find that the {\it w}-mode amplitude near the source would be of 
order $h_+^{(0)} \sim 10^{-9}$. Using this amplitude, a frequency of 10 kHz in the {\it w}-mode
range $8-16$ kHz, we can determine the flux of the gravitational wave near the source 
({\it i.e.} at $r^{(0)}=3 \times 10^4$ m) using the formula \cite{Schutz96}
\begin{equation}
    \label{detect-2}
    F^{(0)} _{gw} = \frac{c^3}{16 \pi G} |{\dot \epsilon}|^2 = 
    \left( 3 \times 10 ^{35} \frac{W s^2}{m^2} \right) h_+^2 f^2 =
    3 \times 10^{25} \frac{W}{m^2} ~,
\end{equation}
where $\epsilon = h_+ e^{iku}$ as defined below equation \eqref{GWmetric}. We can now calculate
the flux of the counterpart VLF electromagnetic radiation using \eqref{detect-2} in
\eqref{emgwRatioB} to give 
\begin{equation}
    \label{detect-3}
    F^{(0)} _{em} = 2 \times (10^{-9})^2 \times (3 \times 10^{25} \frac{W}{m^2})
    = 6.0 \times 10^7 \frac{W}{m^2}~.
\end{equation}
From \eqref{detect-2} and \eqref{detect-3} we find that $F^{(0)} _{gw} \gg F^{(0)} _{em}$
which is expected -- the electromagnetic radiation produced is much smaller than the
gravitational wave which produced it. However, $F^{(0)} _{em}$ is nevertheless large enough
that even taking into account the $1/r$ fall off one could potentially detect this
electromagnetic radiation at the location of the Earth's orbit. Taking
the flux $F^{(0)} _{em}$ from \eqref{detect-3} one can determine the flux at the 
location of the Earth assuming that the neutron star is 1 kpc away.
\begin{equation}
    \label{detect-4}
    F_{em}= F^{(0)} _{em} \left( \frac{r^{(0)}}{1 kpc}\right) \sim 
    6.0 \times 10 ^{-23} \frac{W}{m^2} ~,
\end{equation}
where $r^{(0)} = 3 \times 10 ^4$ m from before. A flux of the magnitude in
\eqref{detect-4} could be detected \cite{Jones17} and given the frequency range
of the {\it w}-modes the associated VLF electromagnetic radiation would 
have a
frequency that is above the plasma cutoff at the location of Earth's
orbit as given in Fig. \eqref{CutoffsFig}. Thus the proposal to detect such
the hypothesized co-produced VLF electromagnetic radiation, coming {\it w}-modes of
neutron star quakes, would be to place a satellite capable of detecting such
radiation near earth's orbit \cite{Jones17, Gretarsson18}. The old Explorer 49
satellite was capable of detecting such VLF electromagnetic radiation. The Explorer
49 satellite was a Lunar orbiting satellite which was periodically occulted
by the Moon in order to block out interference from Solar emissions. The
occultation allows one to detect weak signals like \eqref{detect-4} above
the interference from the Sun. 
            
\section{Conclusion and Future prospects}

For over 100 years there was nothing to support Faraday's expectation of a relationship between 
gravity and electromagnetism. However, in the past 50 years we have seen the development of considerable 
theoretical support for this relationship and in particular the relation between gravitational and 
electromagnetic radiation. The earliest work was by Gertsenshtein \cite{Gertsenshtein60} 
demonstrating that electromagnetic radiation can produce gravitational radiation. This 
was followed in 1975 by Skobelev \cite{Skobelev75} who calculated the small but non-zero 
amplitudes for the $graviton + graviton \to photon + photon$ processes. Beginning around the same time as reference \cite{Skobelev75}, there was work that 
examined the production of electromagnetic fields/multiple photons from a gravitational background \cite{unruh-det,dewitt-det,hawking,unruh,Jones15}. Rrcently we have worked on calculations of the production of electromagnetic radiation 
by gravitational waves propagating in vacuum \cite{Jones16,Jones17,Jones18}. Perhaps 
in the next 50 years we will see empirical evidence  of the relation between 
gravitational and electromagnetic radiation by either direct or indirect observation.

The most promising possibility for direct observation is the detection of VLF counterpart 
production by gravitational waves from neutron star quakes  \cite{Jones17}. This would 
only be possible using space based detectors such as the Voyager missions 
\cite{Kurth84,Kurth03,Webber09} or with a lunar occulted detector similar to the 
Explorer 49 mission  \cite{Gretarsson18}. However, detection of counterpart production 
locally would be limited to the highest frequencies of the counterpart production 
from neutron star gravitational waves. Detectors in the outer heliosphere would be 
much more effective and rather remarkably the Voyager space craft are still making 
observations in the $2-4~\rm{kHz}$ range of expected counterpart production \cite{Gurnett15}.

Even without direct observation it is possible that the counterpart production of electromagnetic radiation would have important applications in astrophysical processes. One intriguing possibility is in the energetics of core collapse supernovae. The prompt production of gravitational waves from the core collapse would produce gravitational waves with quadruple amplitudes on the order of $1~\rm{m}$ and strain amplitude of something like $10^{-5} - 10^{-4}$ in the star layers just outside the core. These strain amplitudes have the potential of producing counterpart radiation of sufficient energy to contribute to the energetics of the supernovae. Previous work on fully general relativistic magnetohydrodynamics \cite{Font07} (MHD) have assumed the ``ideal" MHD condition. This assumption suppresses any potential production of electromagnetic radiation from the strong gravitational wave background. More recent work \cite{Obergaulinger14,Just18,Obergaulinger18} on the effects of magnetic fields and rotation on the energetics of core collapse supernovae have not been fully general relativistic and again could not include the energy from production of electromagnetic radiation by the outgoing gravitational wave.  Fully general relativistic MHD simulations have been implemented \cite{Lehner12_86} for collapsing hyper-massive neutron stars but not for the study of core collapse supernovae. It is possible that the energy associated with electromagnetic  production by gravitational waves outside the iron core could contribute importantly to the supernova, but only fully general relativistic MHD simulations would account for this phenomena in the processes of core collapse and explosion.

Since the production of electromagnetic radiation by gravitational waves is so fundamental it is likely that further study of this phenomena could illuminate our understanding of nature. One recent example of the potential importance of production of photons in a gravitational wave background is the investigation of graviton-photon oscillations in alternatives to general relativity \cite{Cembranos18, ejlli}. This investigation did not directly study counterpart production and potential general relativity violations but does describe the significance of this phenomena in investigating theories of gravity. Counterpart production by gravitational waves \cite{Ricciardone17} could also be important in studies of cosmology. Following the Planck epoch the Standard Model fields were still massless
for some time. It would be interesting to consider the production of the massless Standard Model particles by primordial gravitational waves during the grand unification epoch and prior to the Standard Model particles acquiring mass via the Higgs mechanism.

\end{document}